\documentclass[fleqn,10pt]{wlscirep}
\usepackage[utf8]{inputenc}
\usepackage[T1]{fontenc}
\usepackage{graphicx}
\usepackage{dcolumn}
\usepackage{hyperref}
\usepackage[mathlines]{lineno}
\usepackage{bbold}
\usepackage{tabularx}
\usepackage{cleveref}
\usepackage{physics}
\usepackage{siunitx}
\usepackage{multirow}
\usepackage{subcaption}
\title{Markovian Noise Modelling and Parameter Extraction Framework for Quantum Devices}

\author[1,*]{Dean Brand}
\author[2,3]{Ilya Sinayskiy}
\author[1,3]{Francesco Petruccione}
\affil[1]{School of Data Science and Computational Thinking and Department of Physics, Stellenbosch University, Stellenbosch 7602, South Africa}
\affil[2]{School of Chemistry and Physics, University of KwaZulu-Natal, Durban 4001, South Africa}
\affil[3]{National Institute for Theoretical and Computational Sciences (NITheCS), South Africa}

\affil[*]{deanbrand@proton.me}

\begin{abstract}
In recent years, Noisy Intermediate Scale Quantum (NISQ) computers have been widely used as a test bed for quantum dynamics. This work provides a new hardware-agnostic framework for modelling the Markovian noise and dynamics of quantum systems in benchmark procedures used to evaluate device performance. As an accessible example, the application and performance of this framework is demonstrated on IBM Quantum computers. This framework serves to extract multiple calibration parameters simultaneously through a simplified process which is more reliable than previously studied calibration experiments and tomographic procedures. Additionally, this method allows for real-time calibration of several hardware parameters of a quantum computer within a comprehensive procedure, providing quantitative insight into the performance of each device to be accounted for in future quantum circuits. The framework proposed here has the additional benefit of highlighting the consistency among qubit pairs when extracting parameters, which leads to a less computationally expensive calibration process than evaluating the entire device at once.
\end{abstract}
\begin{document}

\flushbottom
\maketitle
\thispagestyle{empty}

\section*{Introduction}
\label{sec:Introduction}

The engineering of quantum computers is based on a fundamental set of conditions, such as DiVincenzo's criteria \cite{DiVincenzo2000}. These criteria essentially encompass the ability to create, manipulate, and measure a collection of quantum states. These procedures necessarily require the presence of the quantum states in a surrounding environment. When in contact with such an environment, with many additional degrees of freedom for interaction, the theory of open quantum systems (OQS) is required to describe the dynamics of the quantum states involved \cite{Breuer2007}. This framework provides insight into the mechanisms and behaviours of interacting quantum states that dissipate energy and lose coherence within quantum devices. These mechanisms of dissipation are forms of quantum noise, which is a significant hindrance to the improvement of modern quantum technologies, known as NISQ devices \cite{Preskill2018}.

The open system dynamics are described by a quantum master equation which is typically an extension of simple unitary evolution of a quantum system to include dissipative effects introduced by the environment. Solving the master equation for a particular system provides the quantum channel, in the form of a dynamical map, which is a conduit for the evolution of the system based on all the present influences \cite{Nielsen2010}. In the most general form, master equations include too many considerations to be easily solved for all scenarios. However, with reasonable simplifying assumptions, such as the Born-Markov approximation which treats the system as memoryless, a Markovian master equation in the Gorini-Kossakowski-Sudarshan-Lindblad (GKSL) form can be used to model many quantum devices \cite{Gorini1975,Lindblad1976}.

Ideal noisy quantum devices can be described by Markovian models and can be accurately modelled by the GKSL master equation since the energy they dissipate will leave the system. This is contrary to non-Markovian dynamics, in which the dissipated energy can return to the system as a time-correlated excitation with new noise influencing the system, which makes the dynamics much more difficult to predict. Despite the distinction between regimes being very influential on the performance of quantum technologies, it is not often a focal point in discussions of the capabilities of the devices.

One such example is the IBMQ set of quantum devices, openly accessible through the IBM Quantum Experience platform \cite{IBMQ2023}, which has been the test bed of a lot of quantum research on quantum computation and OQS models. These devices serve as a prime example of the current state of NISQ technologies, and as such have been used in support of competing claims of the compatibility of Markovian descriptions with the dynamics present in these devices under various conditions. Like all modern superconducting qubit devices, the primary decay mechanisms that limit performance are relaxation and decoherence, typically characterised by $T_1$ and $T_2$, respectively, as characteristic time scales of coherence. These calibration benchmarks, among others, have become standard measurements of quantum computing performance, as reflected in the amount of research published on the topic in recent years \cite{Klimov2018,Schloer2019,Burnett2019,Carroll2022,DeGraaf2020,Vepsaelaeinen2020}.

Although much research has been done on all of these topics separately, there is a lack of research on the overlap of them all, particularly in the realm of application to currently operational NISQ devices.

Much of the literature investigating areas near this topic has been focused on the presumption that modern superconducting quantum devices behave predominantly according to non-Markovian dynamics\cite{Chen2020,Cerrillo2014,Xiang2021,Pollock2018,White2020}. This is in contrast to the ideal noise models that manufacturers aim for in producing high-quality NISQ devices, which are designed to be shielded from non-Markovian decoherence. For example, the work by Arute et al. \cite{Arute2019} in Google Quantum AI's demonstration of quantum advantage on a superconducting NISQ device, only simple Pauli errors and localized noise models are used, implying purely Markovian dynamics. This contrast between manufacturer claims and published scientific literature brings to light the question of which claims are correct and necessitates an explanation for the disagreement. In the aforementioned literature, such as the work of Pollock et al. \cite{Pollock2018,Pollock2018a}, and White et al. \cite{White2020,White2021a,White2021}, there is a basis of observation that superconducting quantum devices are distinctly non-Markovian, and methods are proposed for how to deal with this kind of noise in quantum computers. While these are significant contributions to the research and development of NISQ devices, there is not sufficient explanation for the deviation from a Markov model to describe the qubit dynamics. This creates a void in current research which this paper is intended to address, following the inspiration of Lindblad tomography by Samach et al. \cite{Samach2021}, by applying Markovian models to benchmarking algorithms in modern NISQ devices to evaluate the various claims about their dynamics.

The first point of this research is the characterisation of benchmarking metrics, such as $T_1$ and $T_2$ times, in the Markovian or non-Markovian regimes to verify the applicability of simplifying assumptions used in the use of NISQ devices as test beds for quantum dynamics. At first these noise models may seem too simple to be relevant, and although we are aware of the recent research on topics such as cross-talk noise \cite{Patterson2019, Sarovar2020, AshSaki2021, Rudinger2021, Zhao2022, Wang2022}, the investigations focused on in this work are only on local noise induced errors. The results of Arute et al. \cite{Arute2019} support this by demonstrating how only local noise models are necessary for this kind of tomography. The methods used here are tested on IBMQ devices due to their accessibility and commonality with the previously mentioned studies. It is important to note, however, that the framework produced in this work is hardware-agnostic, and can be used to characterise any qubit-based quantum system. The second point of this research is a framework which we offer as a new method for real-time calibration of several hardware parameters within one comprehensive procedure, which provides quantitative insight into the performance of each device to be accounted for in future quantum circuits. We also offer insight into the necessity of broad-scale tomography of these devices by showing situations in which a series of smaller tomography procedures can be equally effective and more efficient than the typical computationally expensive standard full tomography procedures.

This paper is structured as follows. In Background Theory we discuss the necessary theory behind the fields that overlap in this work. In Framework we introduce the parameter extraction framework. In Demonstration Procedure the methodology of this demonstration is outlined. In Analysis and Discussion the results are discussed in greater detail along with their implications. In Conclusion we summarise and provide concluding remarks.

\section*{Background Theory}
\label{sec:Background_Theory}

In the case of this work, the quantum devices being used for demonstration are IBM's superconducting transmon qubit processors, which operate through the use of super-cooled superconductors making use of artificial atoms comprised of Cooper Pairs (CPs). It is the presence, or lack thereof, of these pairs in the circuits that is the basis of the two-level system (TLS) necessary for a quantum computational basis. The transmon architecture of an anharmonic quantum oscillator introduces the non-linearity between energy levels which ensures that the quantum states can be individually addressed without accidental driving of higher energy levels \cite{Krantz2019,Huang2020}.

\begin{figure*}[t]
	\begin{center}
		\includegraphics[scale = 0.35]{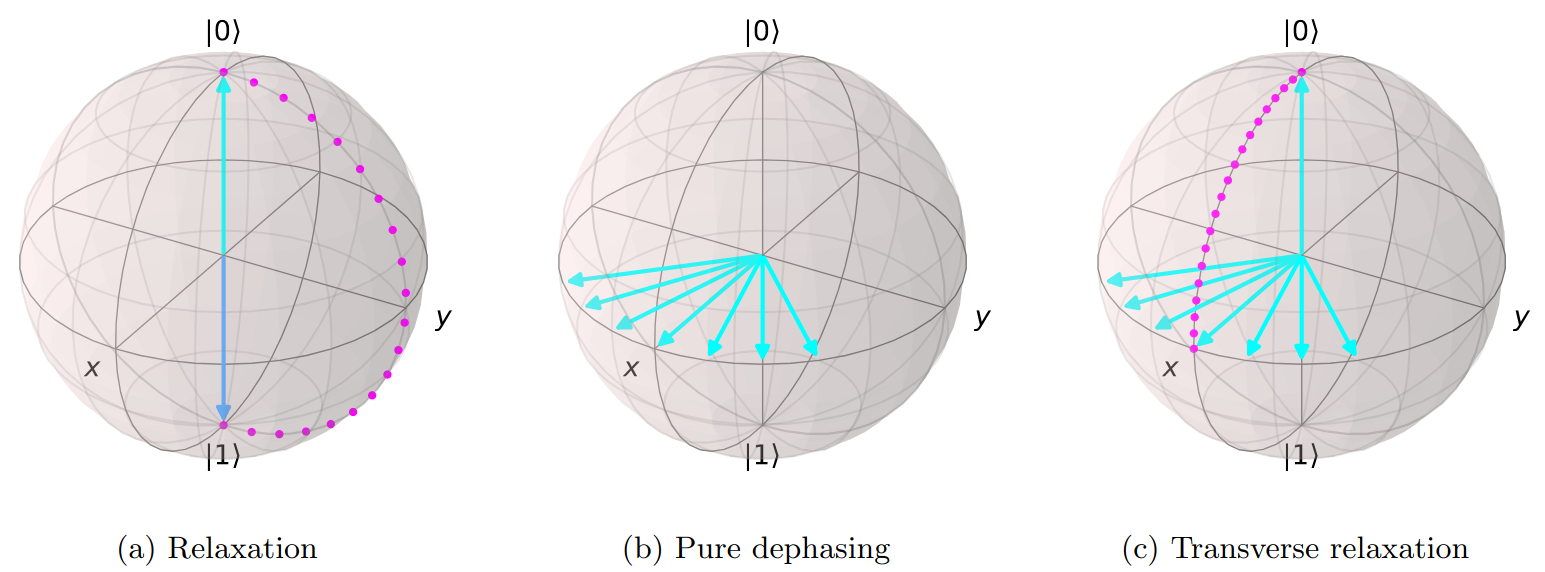}
	\end{center}
	\caption{A Bloch sphere representation of the noise processes of relaxation (a), pure dephasing (b), and transverse relaxation (c), respectively. Relaxation is the process of an excited state, $\ket{1}$, decaying to the ground state, $\ket{0}$. Pure dephasing is the fluctuations along the $x$-$y$ axis from the initial $\ket{+}$ state in this example. Transverse relaxation is the combination of starting in the ground state, moving to the transverse axis, dephasing, and going back to the ground state.}
	\label{fig:bloch_noise}
\end{figure*}

To allow for the full potential of quantum computation to be realised, these qubits can be used not only individually, but in conjunction with each other for a full ensemble where the exponential performance increase can be obtained. This requires the ability of the qubits to connect to each other in the processor to create entanglement and superposition.

The energy of the transmon qubits manifests in the form of qubit frequency, which is a parameter that can be directly controlled and maintained to a high degree of accuracy which assists in the individual addressing of the qubits. This frequency is typically on the order of $5\,\si{\giga\hertz}$. Additionally, to allow for the use of superconducting phenomena, the circuits need to be below the critical temperature at which the materials used become superconducting. In the case of most transmon devices, the processors are kept at temperatures on the order of $10\,\si{\milli\kelvin}$, although this number is not stringently controlled due to the ambient fluctuations in temperature stemming from various sources.

These temperature fluctuations are one of many sources of noise in the system which deteriorates the ability of the quantum states to store information, known as coherence. It is this coherence which allows for quantum computation as without it no information can be manipulated through a computation. Due to the natural fragility of quantum states which are consistently undergoing interactions with the environment, this window of coherence in which computations can be performed is very small. This characteristic time is the coherence time of a qubit, and in the case of NISQ transmon devices is on the scale of $100\,\si{\micro\second}$.

There are two dominant forms of a coherent quantum state decaying in a quantum computer, namely the processes of relaxation and decoherence, characterised by $T_1$ and $T_2$ times, respectively, which are typical performance benchmarks inherited from older realisations of quantum experiments, such as Nuclear Magnetic Resonance (NMR)\cite{Jarek1997}. These processes effectively describe the evolution of an excited quantum state decaying to the ground state through simple spontaneous emission (relaxation) and the transverse coupling to environmental noise to reach equilibrium with the rest of the system (dephasing). These processes have been depicted in a Bloch sphere representation in \Cref{fig:bloch_noise}.

The dynamics of these qubits are typically described only by the system Hamiltonian, which in the case of IBMQ transmon qubits has the form of a Duffing oscillator\cite{Khezri2016}. For example, for a single-qubit device, the Hamiltonian in terms of qubit frequency, $\omega_{q}$, and anharmonicity, $\Delta$, is
\begin{equation}
	\label{eq:single_duff}
	\mathcal{H} = \frac{\omega_{q}}{2}(\mathbb{1}-\sigma^{z})+\frac{\Delta}{2}(O^2-O)+\Omega_{d}D(t)\sigma^{x},
\end{equation}
where $O = b^\dagger b$, $b^\dagger = \sigma^{+}$, $b = \sigma^{-}$, $b^\dagger + b = \sigma^x$ are the operator transformations used and $\Omega_{d}, D(t)$ are qubit-drive parameters. The $\sigma$ operators throughout this work refer to the Pauli matrices. Similarly, but with more detailed structure with included inter-qubit coupling, $J$, the 2-qubit Hamiltonian as defined by the IBMQ backend is
\begin{equation}
	\label{eq:two_duff}
		\mathcal{H} = \sum_{i=0}^{1}\left(\frac{\omega_{q,i}}{2}(\mathbb{1}-\sigma_i^{z})+\frac{\Delta_{i}}{2}(O_i^2-O_i)+\Omega_{d,i}D_i(t)\sigma_i^{x}\right) + J_{0,1}(\sigma_{0}^{+}\sigma_{1}^{-}+\sigma_{0}^{-}\sigma_{1}^{+}).
\end{equation}
This Hamiltonian is projected into the zero excitation subspace of the qubit coupling resonator buses, which leads to an effective qubit-qubit flip-flop interaction. The qubit resonance frequencies in this Hamiltonian are cavity dressed frequencies, while those returned by the backend device include dressing due to qubit-qubit interactions.
It is important to note that these are approximate forms of qubit Hamiltonians, which are claimed by the manufacturers of the devices under the assumption that they describe the dynamics accurately enough for the operation of the device. This is a significant factor which the present work focuses on, to evaluate the validity of this approximation in accurately describing qubit dynamics.

Furthermore, through the use of standard tomography procedures, devices are regularly calibrated to present the most recent parameters of the devices for each qubit, including values of each qubit frequency, anharmonicity, gate error, readout error, and $T_1$ and $T_2$ relaxation and decoherence times. Devices are also kept in dilution refrigerators to maintain the superconducting temperature requirements of the devices, at a claimed temperature of approximately $15\,\si{\milli\kelvin}$, although this is not included in calibration procedures. These are meant as indicators of device performance, with the coherence times and errors being optimised for each new device introduced to the ensemble. These parameters are extracted through simple processes, such as $T_1$ and $T_2$ measurements, which typically do not account for all external sources of noise that influence these dynamics.

\section*{Framework}
\label{sec:Framework}

This work goes into a detailed investigation of the accuracy of the calibrations introduced in the previous section, as well as to investigate additional factors that are not included in the approximate forms of the parameter extractions used in the device calibration. The investigations included here are performed on multiple backend devices across multiple architecture iterations, to investigate which behaviours are not being accounted for in the advancement of the IBMQ family of quantum computers.

\begin{figure}[t]
	\begin{center}
		\includegraphics[scale = 0.8]{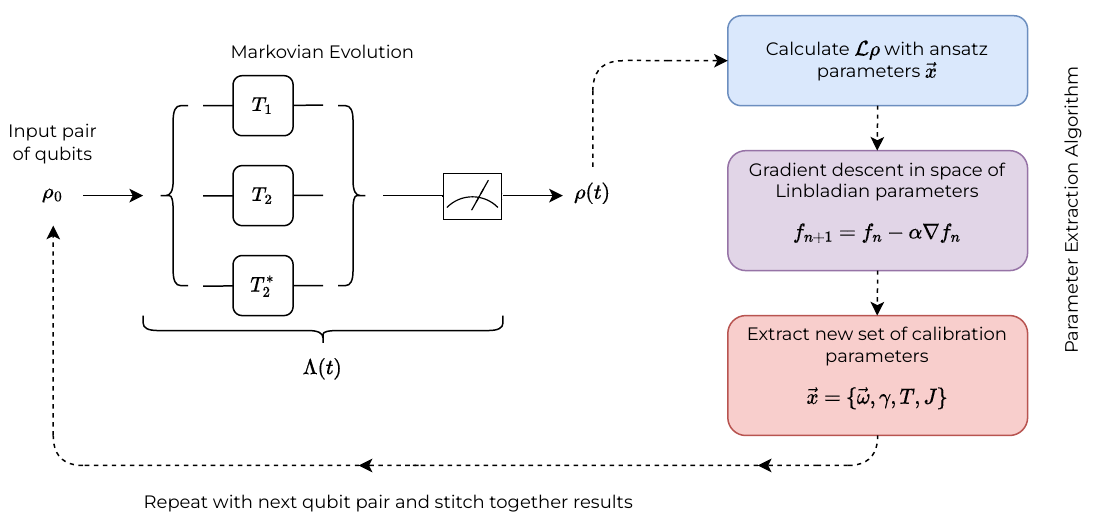}
	\end{center}
	\caption{Noise modelling and parameter extraction algorithm. An input pair of qubits in an initial state $\rho_0$ undergoes Markovian evolution through a benchmark procedure (of $T_1$, $T_2$, or $T_2^*$) allowing for the observation of noise and decoherence in the system. Using hardware parameters the ideal dynamics are calculated through the GKSL master equation. The numerical and observed results are compared, and the hardware calibration parameters are optimised to accurately reflect the system conditions. This process is repeated for qubit pairs throughout the full device to stitch together an accurate description of the quantum device.}
	\label{fig:framework}
\end{figure}

The framework outlined in this section is the basis of the comprehensive noise modelling and parameter extraction process. It is significantly accelerated through the use of automatic differentiation \cite{Baydin2018} in the computational process of numerically solving the relevant equations, to be introduced in this section, and optimising hardware parameters through a gradient descent approach. The methodology of this framework is generalised to model any quantum system susceptible to Markovian noise in an efficient procedure that scales linearly in system size. The outline of this framework is depicted in \Cref{fig:framework}.

In the absence of driving, the Hamiltonians in \eqref{eq:single_duff} and \eqref{eq:two_duff} can be simplified to
\begin{equation}
	\label{eq:single_simple}
	\mathcal{H} = \frac{\omega_0}{2}\sigma^z,
\end{equation}
\begin{equation}
	\label{eq:two_simple}
	\mathcal{H} = \frac{\omega_0}{2}\sigma_0^z + \frac{\omega_1}{2}\sigma_1^z + J_{0,1} \left( \sigma_0^+ \sigma_1^- + \sigma_0^- \sigma_1^+ \right),
\end{equation}
for shallow quantum circuits where the anharmonicity and driving parameters do not have the opportunity to make any significant difference. These Hamiltonians have a clear assumption concerning the axes involved in the definitions of the parameters, in that they are rigid along each axis involved. The qubit frequency is assumed to only have a component along the $z$-axis, and the jump operators of the coupling are assumed to flip the qubit Bloch-orientations perfectly and only along one direct axis of coupling.

A more generalised form of these Hamiltonians needs to be investigated to uncover any underlying deviations from the initial assumptions. For the single-qubit case, this generalised Hamiltonian has the following form,
\begin{equation}
	\label{eq:single_general}
	\mathcal{H} = \frac{1}{2} \left( \vec{\omega} \cdot \vec{\sigma} \right), \text{ where }\, \vec{\omega} = \mqty(\omega_x \\ \omega_y \\ \omega_z) \,\,\text{and}\,\,\, \vec{\sigma} = \mqty(\sigma^x \\ \sigma^y \\ \sigma^z).
\end{equation}
Similarly, for the two-qubit case, there are terms for each qubit, denoted by subscripts, and can be expressed as
\begin{equation}
	\label{eq:two_general}
	\mathcal{H} = \frac{1}{2} \left( \vec{\omega}_0 \cdot \vec{\sigma}_0 \right) + \frac{1}{2} \left( \vec{\omega}_1 \cdot \vec{\sigma}_1 \right) + \vec{\sigma}_0 \vb{J} \vec{\sigma}_1,
\end{equation}
where the new coupling matrix term is defined as
\begin{equation}
	\label{eq:coupling_matrix}
	\vec{\sigma}_0 \vb{J} \vec{\sigma}_1 = \mqty(\sigma_0^x & \sigma_0^y & \sigma_0^z) \mqty(J_{xx} & J_{xy} & J_{xz} \\ J_{yx} & J_{yy} & J_{yz} \\ J_{zx} & J_{zy} & J_{zz} \\) \mqty(\sigma_1^x \\ \sigma_1^y \\ \sigma_1^z).
\end{equation}
This is a far more inclusive description of the qubit Hamiltonian, as it will give greater insight into the behaviour of the qubits. This behaviour is not limited to the scalar properties but rather vectorised descriptions which can reveal possible noise sources which are not accounted for.

It is this complex set of dynamics which reinforces the necessity of the theory of open quantum systems to model and predict the evolution of a quantum device performing a calculation or execution of a quantum algorithm. In the framework of OQS, these dynamics are typically encapsulated in a dynamical map, $\Lambda (t)$, which is a collection of completely positive trace-preserving (CPTP) maps, that constructs a quantum channel describing the path of evolution of a collection of interacting quantum states represented by a density matrix, $\rho$. This evolution of the collection of quantum states is easily expressed in terms of a dynamical map as
\begin{equation}
	\label{eq:dynamical_map_evolution}
	\rho(t) = \Lambda(t)\rho(0).
\end{equation}
In many cases of describing quantum devices, the dynamical map will satisfy a time-local master equation,
\begin{equation}
	\label{eq:time_local_master}
	\dv{t}\Lambda(t) = \mathcal{L}\Lambda(t) \Longrightarrow \dv{t}\rho(t)=\mathcal{L}\rho(t).
\end{equation}
This introduces the Lindbladian generator, $\mathcal{L}$, which has the form of
\begin{equation}
	\label{eq:time_local_generator}
		\mathcal{L} \rho(t) = -i \left[ \mathcal{H}, \rho(t) \right] + \sum_{\alpha} \gamma_\alpha \left( A_\alpha \rho(t) A_\alpha^\dagger - \frac{1}{2} \left\{ A_\alpha^\dagger A_\alpha, \rho(t) \right\} \right),
\end{equation}
where the first term describes the unitary evolution of the system in natural units, the coefficients $\gamma_\alpha$ represent the decay rates of the system, and $A_\alpha$ represents noise (or "jump") operators. The form of \eqref{eq:time_local_generator} is the GKSL generator corresponding to the master equation which describes Markovian dynamics, which neglects memory effects of the system and assumes that any information dissipated from the system is not time-correlated with external noise entering the system to influence future dynamics. There are many descriptions of non-Markovian dynamics due to its much broader scope, however in this work the simple definition we will assume is that non-Markovian dynamics are any that are not accurately described by the GKSL master equation.

This Markovian framework allows for the extraction and verification of claimed qubit parameters from the calibration metrics, such as the relaxation and decoherence times, as well as an extraction of information not included in the calibration, such as the qubit temperatures. These capabilities are all included in the GKSL master equation, which for this application is expressed as
\begin{equation}
    \label{eq:n_gksl}
    \begin{aligned}
        \dv{t} \rho = &-i\left[ \mathcal{H}, \rho \right] + \sum_{i = 0}^{N - 1} \bigg[ \gamma_i\left( \expval{n_i} + 1 \right) \left( \sigma_i^- \rho \sigma_i^+ - \frac{1}{2} \left\{ \sigma_i^+ \sigma_i^-, \rho \right\} \right) \\
        &+ \gamma_i\expval{n_i} \left( \sigma_i^+ \rho \sigma_i^- - \frac{1}{2} \left\{ \sigma_i^- \sigma_i^+, \rho \right\} \right) + \gamma_{z,i} \left( \sigma_i^z \rho \sigma_i^z - \rho \right) \bigg].
    \end{aligned}
\end{equation}
where $\gamma$ is the emission coefficient, and acts as an inverse of the relaxation and decoherence times, and $\expval{n}$ represents the average number of photons emitted as the density matrix evolves, which is represented by
\begin{equation}
	\label{eq:avg_photon_non_natural}
	\expval{n} = \frac{1}{e^{{\hbar\omega}/{k_B T}} - 1}.
\end{equation}
It is worth noting that the average photon emission is time independent due to the Markovianity of the system, whereby the emissions of the qubits are not significant enough to alter the reservoir of the device enough for that to influence the qubit over time. The devices are kept in a controlled state of temperature and electromagnetic field strength such that qubit dissipations can be quickly removed from the system.\cite{Rivas2014} The emission coefficient $\gamma_z$ represents the process of pure dephasing (\Cref{fig:bloch_noise}). This encapsulates the ideal Markovian dynamics of the system, even at absolute zero, and allows for the extraction of decay times through the emission coefficient and temperature through the photon number. It is important to note that the explicit inclusion of a pure dephasing contribution to the density operator evolution was considered in additional experiments to quantify its influence. These additional experiments revealed that the presence of pure dephasing was insignificant in the modelling of the qubit systems. This is, however, not enough evidence to suggest that the process of pure dephasing is entirely irrelevant to these quantum devices, rather that it is inconsequential in this method of modelling the qubit evolution and extracting hardware parameters. As such, the quantitative mention of this parameter is omitted from the present discussion.

The present work focuses on a handful of simple conventional quantum noise channels, as well as modifications to these, which are typically used to describe decay times $T_1$, $T_2$, and $T_2^*$. It should be noted that the primary focus of this work is based on 2-qubit demonstrations, and so only the 2-qubit circuit representations will be discussed.

\begin{figure*}[t]
	\begin{center}
		\includegraphics[scale = 0.35]{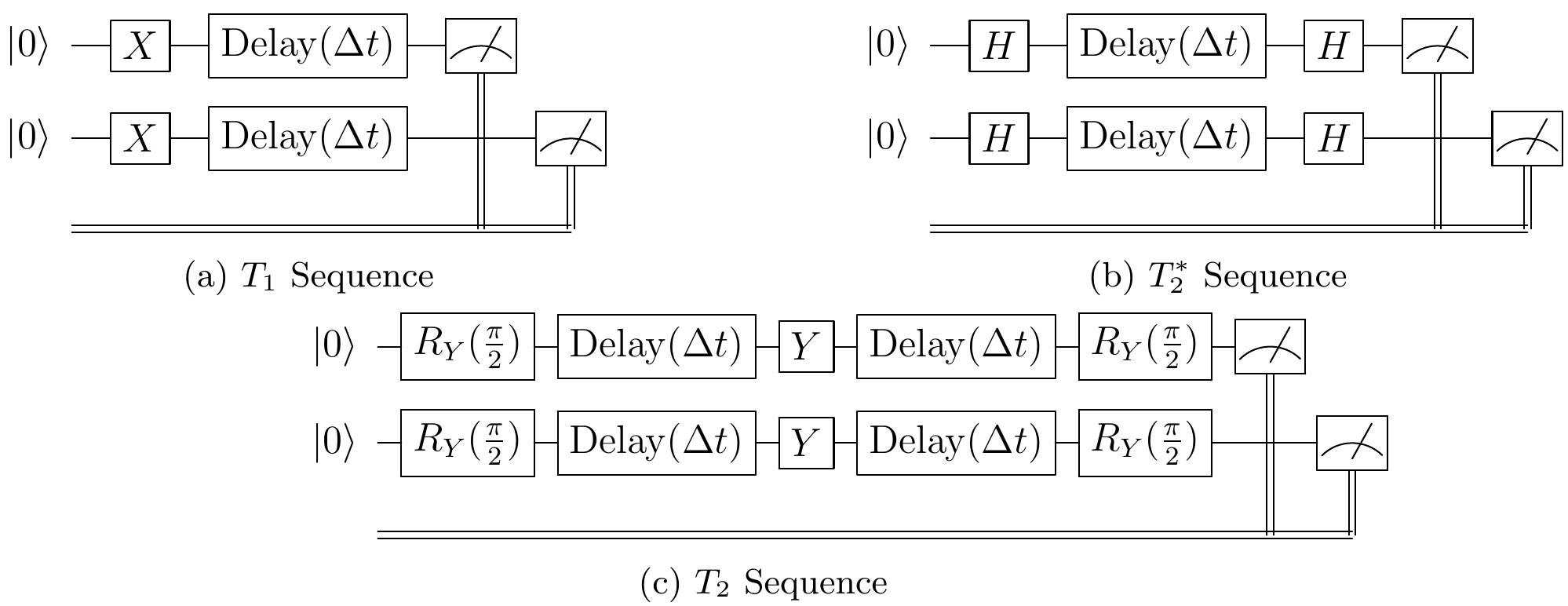}
	\end{center}
	\caption{Benchmark procedure circuit diagrams for 2-qubit systems. Circuits shown for $T_1$ (a), $T_2^*$ (b), and $T_2$ (c) sequences. These circuits are constructed to measure various benchmark data, such as characteristic timescales of noise interference. The control gates, $X$, $H$, and $R_Y$ initialise the desired quantum states, and the delay gates allow for the decoherence and dephasing of the quantum states to occur.}
	\label{fig:basic_circuits}
\end{figure*}

The circuit built to measure the $T_1$ relaxation time is the simplest of the sequences, as depicted in \Cref{fig:basic_circuits}a, and consists in this case of a qubit, or set of $n$ qubits, initialised in the $\ket{0}^n$ state, after which a set of $X^n$ gates is applied to excite the qubits to the $\ket{1}^n$ state. After this excitation, the qubits decay for a variable time, after which the states are measured in the computational basis and the state distribution calculated.

\begin{figure*}[t]
	\begin{center}
		\includegraphics[scale = 0.35]{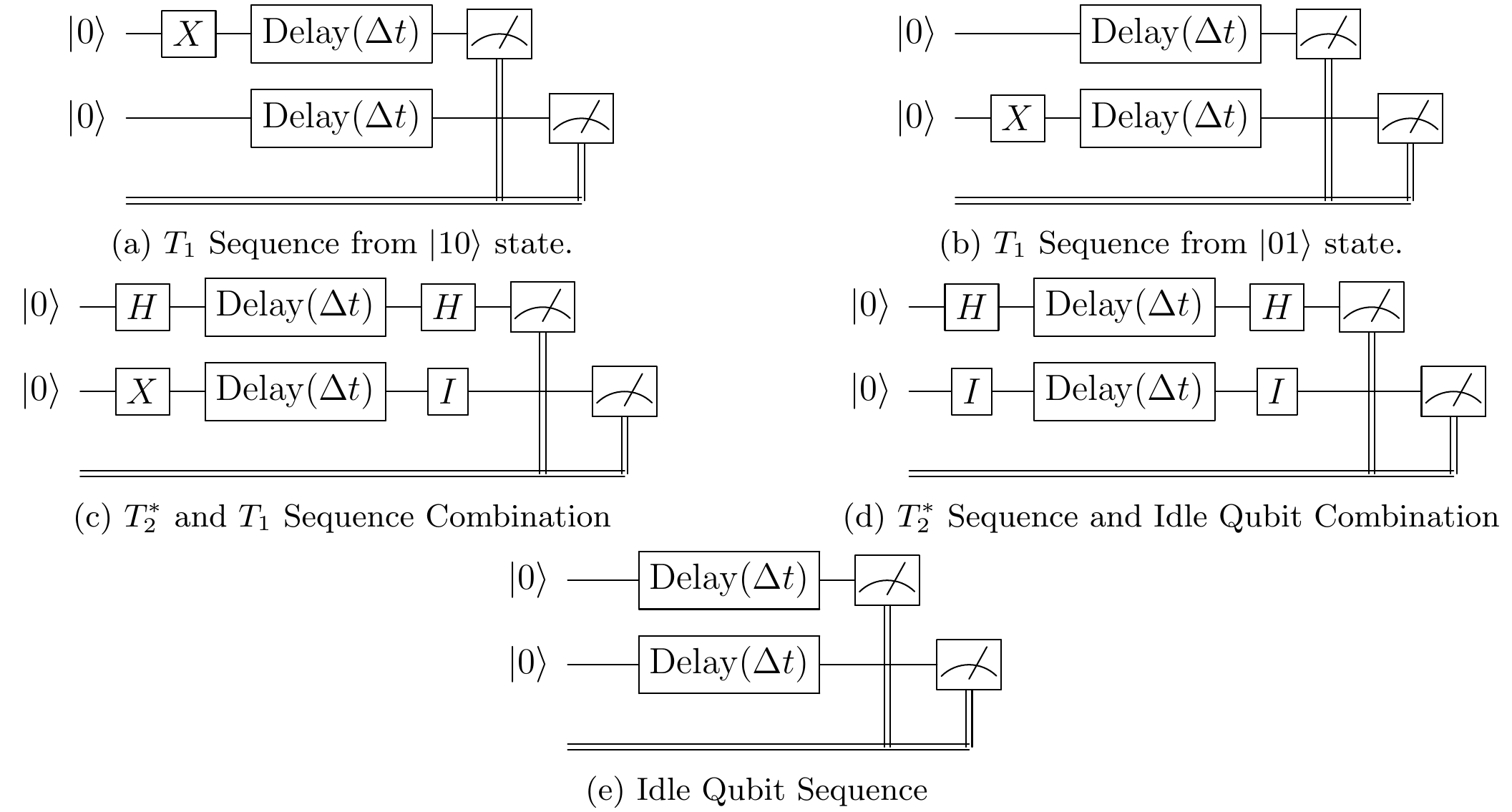}
	\end{center}
	\caption{Combination procedure circuit diagrams for 2-qubit systems. Circuits shown for $T_1$ from $\ket{10}$ (a), $T_1$ from $\ket{01}$ (b), $T_2^*$ and $T_1$ (c), $T_2^*$ and idle (d), and 2-qubit idle (e) sequences. These circuits are constructed to measure various benchmark data, such as characteristic timescales of noise interference. The control gates, $X$ and $H$ initialise the desired quantum states, and the delay gates allow for the decoherence and dephasing of the quantum states to occur.}
	\label{fig:mixed_circuits}
\end{figure*}

The $T_2^*$ procedure, shown in \Cref{fig:basic_circuits}b, consists of the qubit set again being initialised as $\ket{0}^n$, after which a $\pi/2$ rotation is applied, in this case through the Hadamard gate, $H$, to the $\ket{+}$ state. After this, the state is left to undergo pure dephasing for a variable time period, after which another Hadamard operator is applied to return the qubits to the $\ket{0}$ state where they can be measured.

For the procedure $T_2$, the Hahn echo sequence \cite{Hahn1950} is used, shown in \Cref{fig:basic_circuits}c, which consists of a set of $n$ qubits initialised in the state $\ket{0}^n$, after which a set of $\pi/2$ rotations is applied around the $x$- or $y$-axis. These $\pi/2$ rotations are applied through an $R_Y$ gate, defined as
\begin{equation}
	\label{eq:ry_gate}
	R_Y(\theta) = \exp \left( -i\frac{\theta}{2}Y \right) = \begin{pmatrix}
		\cos\frac{\theta}{2} & -\sin\frac{\theta}{2} \\
		\sin\frac{\theta}{2} & \cos\frac{\theta}{2} \\
	\end{pmatrix}.
\end{equation}
Succeeding this is a delay period of variable time followed by a $\pi$ rotation (as $Y = R_Y(\pi)$) around the same axis as the first rotation, and then another delay time of the same period, and a final $\pi/2$ rotation in the same direction to return the state to $\ket{0}^n$ where it can be measured.

The choice of $T_1$, $T_2$, and $T_2^*$ as benchmark demonstrations has the additional benefit of being measurements of the dynamics of $\expval{\sigma^x}$, $\expval{\sigma^y}$, and $\expval{\sigma^z}$, respectively. These procedures are extended uniformly across the various qubit sizes involved in the demonstrations, namely ensembles of 1, 2, and 3 neighbouring qubits. The fact that the investigated qubits are neighbours means that their direct coupling strength may lead to stronger interference between them. Though this phenomenon is necessary for multiqubit gates, which is a reason why they are coupled at all, there may be inadvertent effects of this coupling creeping into isolated channels of individual qubits. To further investigate this possibility, we adapted the following two-qubit sequences from the original procedures described above.

The first modified sequences are made by changing the $T_1$ procedure, which was initialised as $\ket{00}$, and excited by simultaneous $X$ gates before the relaxation period. This procedure has the qubits mimic each other's effects, which makes it difficult to detect the presence of coupling phenomena between them as their behaviour is indistinguishable from a scenario wherein they are coupled. This is circumvented by performing the demonstrations not only with simultaneous excitation but also with quantum circuits wherein only one of the qubits is excited through an $X$ gate and the other is left to evolve from the $\ket{0}$ state, as in \Cref{fig:mixed_circuits}a and \Cref{fig:mixed_circuits}b.

If the coupling is strong enough and not shielded in some way, then the decay of the excited qubit will directly influence the stationariness of the other. By performing the demonstration with the exclusion of excitation operators entirely, shown in \Cref{fig:mixed_circuits}e, the qubits are left in the ground state to idle for the delay period to check if there are any external sources which unintentionally excite the subsystem of qubits.

To provide further insight into the relaxation and decoherence mechanisms of the qubits, the procedures described before can be combined into new composite systems to show the dynamics between various decay mechanisms, and how the decay of one qubit in the system might influence its neighbours which are ideally excluded from the subsystem. To achieve these, the 2-qubit ensemble is modified into new quantum circuits, the first of which has one qubit undergoing a standard $T_2^*$ procedure, while its neighbour undergoes a $T_1$ procedure, illustrated in \Cref{fig:mixed_circuits}c. Similarly, the other modified circuit has the first qubit undergo a sequence $T_2^*$, while the neighbour qubit simply stays in its idle state from the initial state $\ket{0}$, as shown in \Cref{fig:mixed_circuits}d.

To elucidate what should happen from a theoretical point of view, it is necessary to return to the master equation \eqref{eq:n_gksl}. This equation is complicated to solve analytically, so numerical methods are typically used to obtain useful information. A solution to this equation can be found for a set of parameters, $\vec{x}$, which depends on the form of the master equation and the Hamiltonian used. For example, the single-qubit master equation solution with the simple Hamiltonian \eqref{eq:single_simple} is a function of 4 parameters,
\begin{equation}
	\label{eq:single_simple_params}
	\vec{x} = \left( t, \omega, \gamma, T \right),
\end{equation}
being time $t$, qubit frequency $\omega$, emission rate $\gamma$, and temperature $T$. However, for the general Hamiltonian \eqref{eq:single_general}, the solution is a function of 6 parameters,
\begin{equation}
	\label{eq:single_general_params}
	\vec{x} = \left( t, \omega_x, \omega_y, \omega_z, \gamma, T \right).
\end{equation}
The size of the parameter vector quickly grows for multi-qubit states, as in the example of a 2-qubit subsystem, the solution will require 8 parameters for the simple Hamiltonian \eqref{eq:two_simple}, and 20 parameters for the generalised Hamiltonian \eqref{eq:two_general}. Nevertheless, these equations are numerically solvable as functions of these parameters and initial states from the qubits at the start of the delay period, to return a time series of the evolution of the density matrix. The solution of the master equation can be expressed as the integral of
\begin{equation}
	\label{eq:master_eq_sol}
	\dv{t}\rho(t)=\mathcal{L}\rho(t) \qquad \rho_0 = \rho(t = 0) = \op{0},
\end{equation}
in the case of the single-qubit sequence $T_1$, where the system starts in the state $\ket{0}$ and is excited to $\ket{1}$. This evolution through the delay periods in the quantum circuits can be combined with the quantum gates as operators in the construction of the quantum channels to describe the entire evolution of the system.

The full solution of the master equation for a set of parameters which is passed through the Kraus form\cite{Rivas2012} of the quantum channel is then a set of values which are comparable to the demonstration results which are obtained through the respective quantum circuit. Through the use of the parameters provided by the periodic device calibration, the master equation solution can be compared directly to the demonstration results to verify the accuracy of the calibration data. This verification process allows for all of the hardware parameters to be verified in conjunction with each other, rather than the independent procedures that were used to extract those values initially.

It is important to note that the discussion of calibration data used here is not crucial to this framework and experimental procedure, which is independent of the device used. The calibration data obtained from the IBMQ interface is only used as a convenient ansatz of initial parameters. The framework is independent of initial parameters, which can be randomised or selected through any method the user prefers. The use of available calibration data does assist in faster convergence to optimised parameters and avoid local optima and exploding/vanishing gradients. It is also worth reiterating that this framework does not obtain the calibration data using the same methodology as the IBMQ backend. The calibration data is obtained through simplified curve fitting for each parameter, while ignoring the other device behaviours. This framework obtains all hardware parameters in one comprehensive procedure, which also has the ability to reflect relations between hardware behaviours which are not seen in simpler models.

Furthermore, this method provides the capability to improve upon the parameter extraction in the case that the claimed parameters do not match the observations. The parameters used in the master equation solution can be iteratively varied to provide different results, until a parameter set is found that accurately matches the hardware data.

This can be achieved very easily through conventional optimisation methods which have shown extreme success in achieving this form of outcome, namely gradient descent. In this work, the more sophisticated and accurate gradient descent method of the adaptive moment estimate (Adam) optimiser \cite{Kingma2017} is used.

The cost function used in this optimisation process is a simple least-squares regression metric, which sums all of the squared values of the difference between each observed and estimated data point,
\begin{equation}
    \label{eq:least_squares}
    S = \sum_{i = 1}^{n} \left( y_i - f_i(\vec{x}\,) \right)^2,
\end{equation}
where $y_i$ are the experimental values.

Using this method to find the minimum difference between the numerical and hardware results yields a set of parameters that more accurately describe the properties of the qubits being operated on. The extraction of these parameters not only allows for a new method of calibrating the device but also gives insight into the way that they are influenced by different experiments and quantum circuits.

\section*{Demonstration Procedure}
\label{sec:Demonstration_Procedure}

Communication with quantum devices is done through the Qiskit SDK \cite{Qiskit}, which allows the extraction of backend configuration information, as well as the construction of quantum circuits that are converted to basis gate circuits to perform the demonstrations. Once these circuits are run, the data can be extracted and used as desired. Qiskit also offers many built-in functionalities such as readout error mitigation and state tomography procedures.

Qiskit allows for a selection of several backend devices to be worked with, and for each, offers a set of calibration data as well as device properties such as the qubit topology. This information was used in this work as the parameters of the numerical solutions to verify the accuracy of the calibrations performed on the backend side. This information also provided scales of extra errors, such as read-out and gate errors, which could be taken into account in analysing the data.

\begin{figure}[t]
	\begin{center}
		\includegraphics[scale = 0.45]{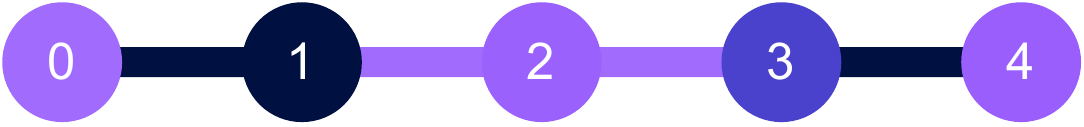}
	\end{center}
	\caption{Qubit topology of 5-qubit \textit{ibmq\_santiago} v1.3.40 and \textit{ibmq\_manila} v1.0.19 devices. The colours of the qubits represent the frequency, with darker meaning a lower frequency and lighter meaning a higher value.}
	\label{fig:Q5_Topology}
\end{figure}

This work made extensive use of the \textit{ibmq\_lima} v1.0.52, \textit{ibmq\_santiago} v1.3.40, and \textit{ibmq\_manila} v1.0.19 devices for multi-qubit circuits as these devices have 5 qubits each. For the results presented in this work, these devices were accessed over a period from November 2022 to July 2023. The topology of these devices (excluding \textit{ibmq\_lima} which has a "T-shaped" layout) is a simple linear structure, as in \Cref{fig:Q5_Topology}, and allows investigation of coupling effects between neighbours.

Once the backend was selected, the circuits could be constructed. The circuits introduced in the last section needed to be slightly modified to obtain useful data. Due to the nature of quantum measurement, the state distribution could not be continuously measured; otherwise, the quantum Zeno effect \cite{Misra1977} would alter the data by not allowing the system to undergo the desired decay. This means that running the circuits as presented, \Cref{fig:basic_circuits} for example, would return only one time-slice of the data and the bit-string probability distribution at that snapshot in time. To avoid this, a series of circuits needed to be constructed and run with the delay time being varied to allow each time step to be a new point in the dataset.

In the execution of these procedures, there are unavoidable errors in the process which detract from the fidelity of the desired hardware results. These errors are primarily SPAM errors\cite{Chen2019,Gambetta2012} in readout and gate execution, which are not important in the present investigation, but plague the data nonetheless. These errors must be taken into account to obtain meaningful data. The gate errors are avoided by being on time-scales small enough for the gate execution times to be insignificant compared to the circuit execution time. As mentioned in previous sections, the probabilistic nature of quantum systems demands that experiments and measurements be made multiple times to create an ensemble with a reliable probability distribution from which average values can be extracted. In this demonstration, each procedure was run for $8\,192$ iterations, or "shots" in Qiskit, which ensures that the variance between iterations is suppressed enough to avoid the influence of gate errors.

The readout errors, however, are present throughout the device and cannot be avoided through collecting more data for each procedure, but rather need to be calibrated for within the hardware run of executing all of the circuits. The measurement calibration functionality built into Qiskit follows a procedure of preparing all of the qubits in the system, or a defined subsystem, in a certain state, which in this case is similar to tomographic methods in that it produces all possible states, and measuring immediately afterwards. This produces a state distribution demonstrating the accuracy of the readout, which can be compared to the ideal case where the outcome from such an process is known analytically. These processes were applied and accounted for in each iteration of the hardware procedures to minimise the influence of SPAM errors throughout the collection of results.

For the calculations of the solution of the master equation and the optimisation algorithm, the JAX SDK \cite{JAXGitHub, JAXPaper} was used as it provides substantial increases in numerical performance, particularly through its auto-differentiation and Just-in-Time (JIT) compiler features, which assist in calculus-based and iterative calculations.

\begin{figure*}[t]
	\begin{center}
		\includegraphics[scale = 0.5]{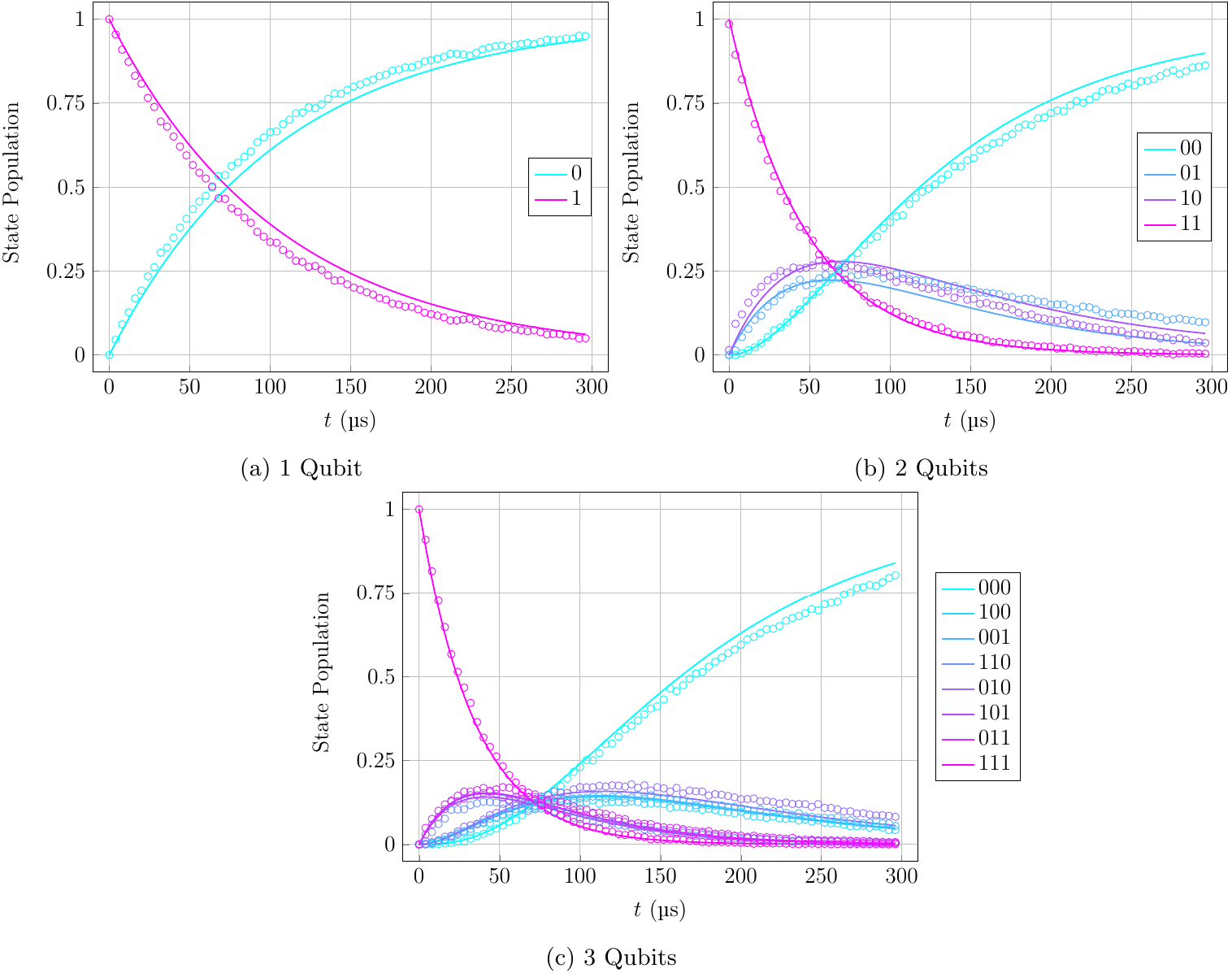}
	\end{center}
	\caption{$T_1$ relaxation density matrix evolution. Hardware data represented by dots and numerical data for claimed hardware parameters represented by solid lines. The $T_1$ time is a characteristic quantum state decay time measuring the rate of decoherence from noise in the system. Results shown for 1-qubit (a), 2-qubit (b), and 3-qubit (c) cases.}
	\label{fig:t1_num_exp}
\end{figure*}

\begin{figure*}[t]
	\begin{center}
		\includegraphics[scale = 0.5]{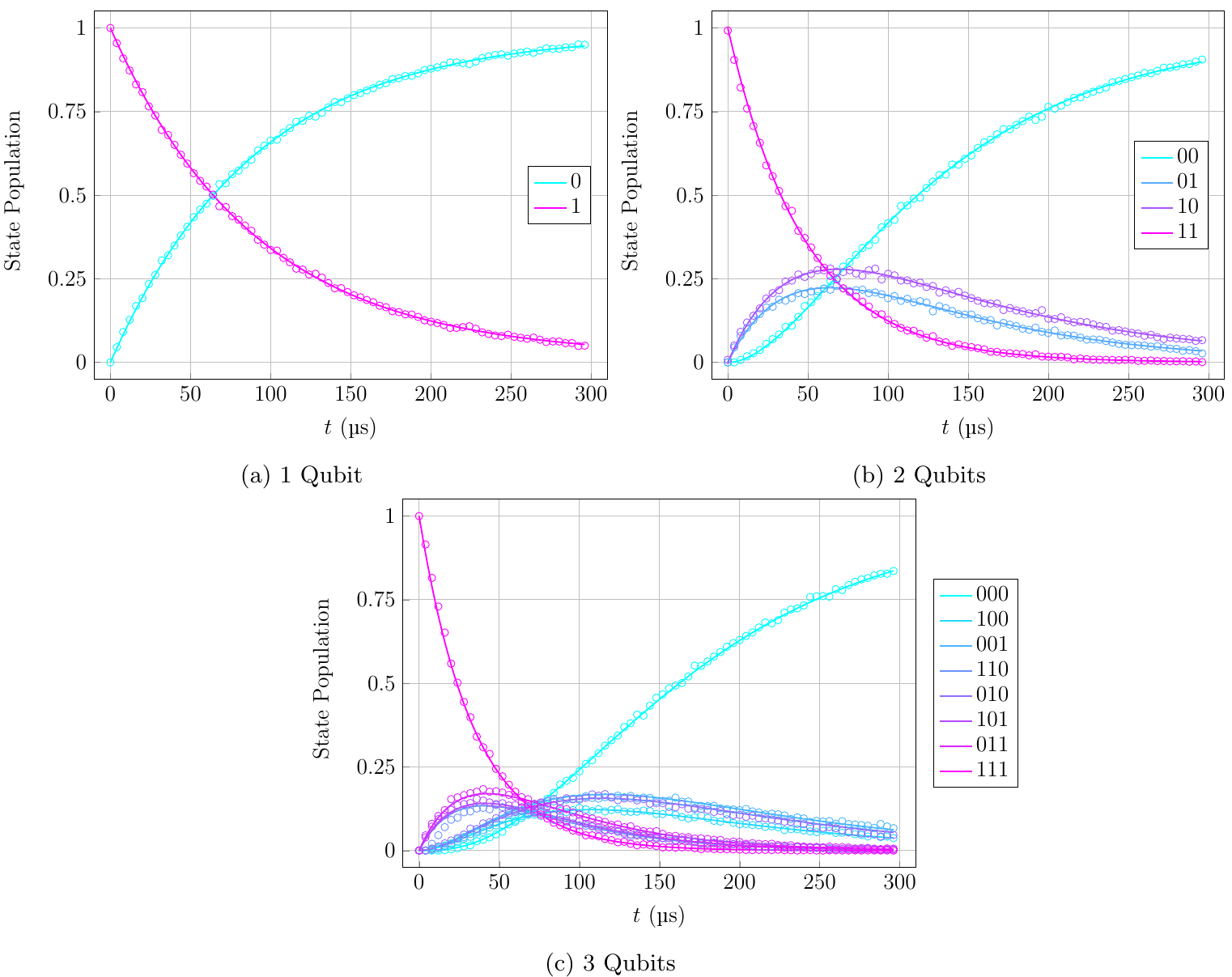}
	\end{center}
	\caption{Optimised $T_1$ relaxation density matrix evolution. The $T_1$ time is a characteristic quantum state decay time measuring the rate of decoherence from noise in the system. Hardware data represented by dots and numerical data for optimised parameters represented by solid lines. Results shown for 1-qubit (a), 2-qubit (b), and 3-qubit (c) cases.}
	\label{fig:t1_opt_fit}
\end{figure*}

\begin{figure*}[t]
	\begin{center}
		\includegraphics[scale = 0.5]{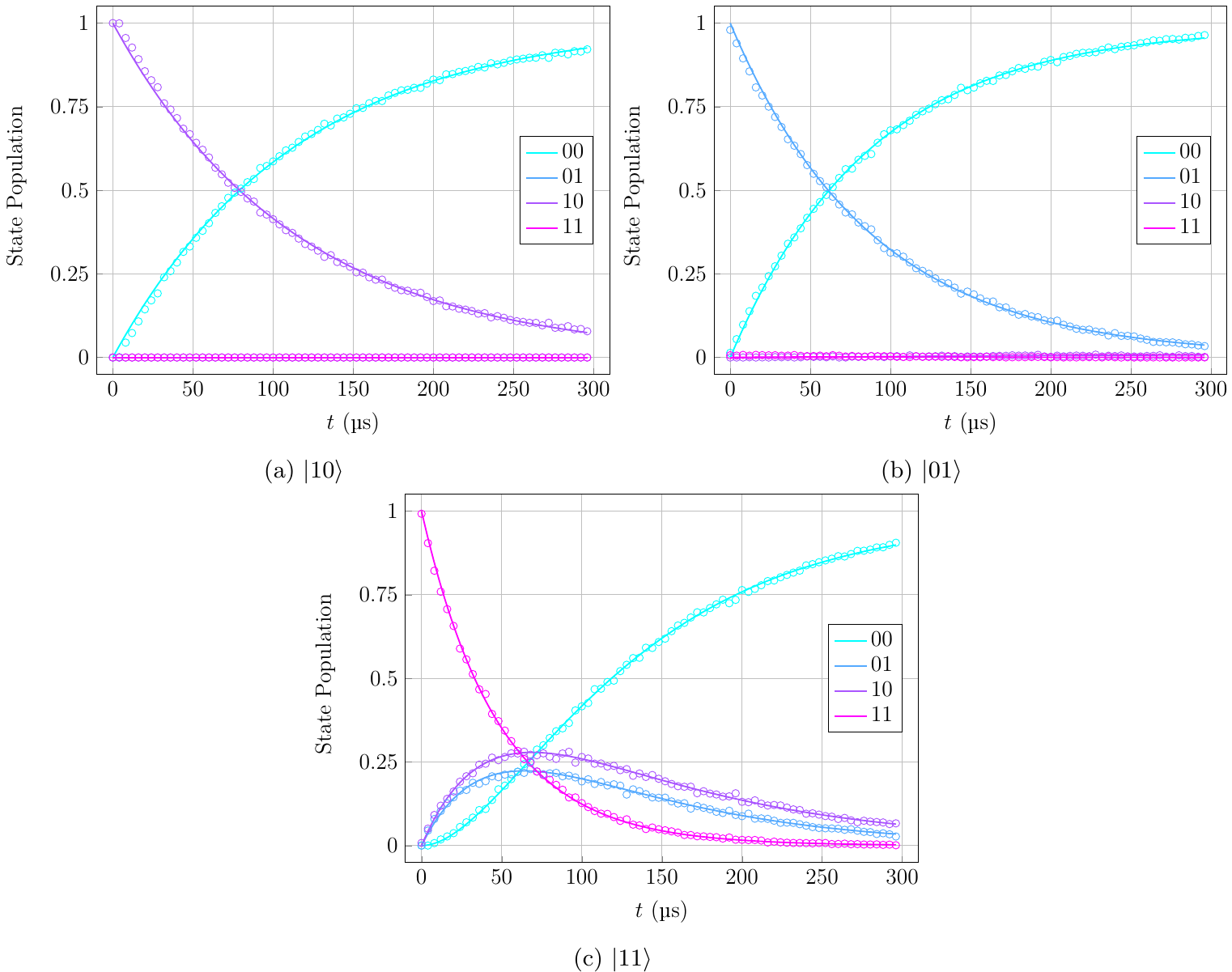}
	\end{center}
	\caption{Optimised $T_1$ relaxation density matrix evolution for varied initial conditions, obtained by circuits shown in \Cref{fig:mixed_circuits}. The $T_1$ time is a characteristic quantum state decay time measuring the rate of decoherence from noise in the system. Hardware data represented by dots and numerical data for optimised parameters represented by solid lines. Results shown for initial states $\ket{10}$ (a), $\ket{01}$ (b), and $\ket{11}$ (c).}
	\label{fig:t1_inits}
\end{figure*}

To justify the use of the optimisation procedure and this section of the work, it should be noted that the numerical results produced by the solution of Equation \eqref{eq:n_gksl} with the hardware calibration data used as initial parameters created a very close fit to the hardware data, as can be seen in \Cref{fig:t1_num_exp}. The overlay of numerical results on hardware data show that the form of the evolution is correct, however not perfectly accurate due to the hardware parameters not being an accurate depiction of the present state of the system at any given moment after recalibration. This reinforces the claim that the qubit dynamics, at least in the case of these benchmark procedures, are indeed Markovian, and that there is a need to tweak the initial parameters by an optimisation procedure.

In order to further investigate the behaviour of the $T_1$ sequence, particularly in terms of the coupling between multiple neighbouring qubits, the modifications of the sequence using different initial states, as in the circuits represented in \Cref{fig:mixed_circuits}a and \Cref{fig:mixed_circuits}b, the same process of circuit compilation, execution, and fitting was followed. The results obtained by these verified the stability and isolated nature of this simple relaxation sequence, as can be seen in \Cref{fig:t1_inits}. The data show expected behaviour of the interqubit coupling not influencing the Markovianity of the system evolution, which is seen in the ground-state qubits staying in the ground state through the full period, while the excited-state population follows a simple exponential path to equilibrium.

Although the $T_1$ sequences for all of the qubit sizes and initial states proved to be very stable and reliable, the rest of the procedures provided more intricate results. For a first example, the $T_2$ sequence for a single qubit proved to be a significantly more difficult numerical procedure, taking far longer to calculate the results despite the simple appearance of exponential decay, as seen in \Cref{fig:extra_plots}a. Despite the difficulty in extracting numerical solutions and finding optimal parameters, the procedure still found great success in achieving these goals. The optimal parameters provided a very accurate fit, similar to the result of $T_1$, having a least squares error at the level of $10^{-2}$ consistently.

\section*{Analysis and Discussion}
\label{sec:Analysis_and_Discussion}

Apart from the $T_2^*$ sequences, which provided lower fidelity data, all of the hardware procedures proved to attain successful and reliable data which in turn provided valuable insight. The data allowed for very efficient optimisation of hardware parameters, allowing for the probing of these values without direct access to their measurements from a relatively simple theoretical model. This section will focus on the detailed analysis of these results and the numerical accuracy of all the data obtained.

In the case of single-qubit procedures, all sequences proved to be highly successful and reliable in extracting information, mostly because of the simple forms of the qubit dynamics. These results didn't have any surprising features, as the observed results had high fidelity and consistency through iterations, while the numerical results were based on a simple form of the GKSL Master Equation which made the computation very quick and easy, especially in the optimisation process which thereby allowed for more iterations to be run and higher accuracy to be achieved. The set of hardware parameters that were extracted was the smallest of all the subsystems, due to the simple Hamiltonian form, which provided useful insight despite the simplicity. The Hamiltonian parameters extracted, being the qubit frequencies along each Bloch sphere axis $\left( \omega_x, \omega_y, \omega_z \right)$, showed that the simplifying assumption given by the backend providers of there only being a $z$-component is not entirely accurate. This was shown by small but significant contributions to the qubit frequency along the $x$- and $y$-axes, with the vector norm of these being roughly equivalent to the claimed hardware parameter, within around $10\,\si{\mega\hertz}$. This would ordinarily not pose any issue for the function of the devices, so the backend claim of the simplified model is effectively correct in most use cases; however, this general form should not be ignored as it can prove to be a significant factor in the scalability of larger devices with potential resonance with neighbouring qubits.

In the case of the 2-qubit subsystems, the data showed a similar pattern, as the simple $T_1$ sequence provided the most accurate results, giving rise to insightful optimised parameters. These parameters showed once more that the simple effective Hamiltonian claimed by the backend provider does not show the full picture, but rather each qubit has contributions to the frequency from the transverse Bloch sphere axes, and a similar pattern is observed in the qubit coupling parameters which are not the simplified scalar values claimed to act only upon the $x$- and $y$- axes, but rather a $3 \times 3$ matrix of coupling along all axes. This discrepancy does not pose any significant influence on the relaxation and decoherence dynamics investigated here, but once again should be accounted for in considering the scalability of the quantum devices to avoid unwanted noise from resonances. Significantly, in terms of the devices' temperatures, which did not have claimed calibration values, but rather a general order of magnitude claim, the average number of photons of the Markovian dynamics demonstrated that these claims are accurate to within $\pm10\,\si{\milli\kelvin}$, proving that this method is a viable approach to calibrating the device and individual qubit temperatures. However, it should be noted that this temperature measure does have the disadvantage of being an inferred value from the photon emission contribution to state decay and, as such, does carry inherent inaccuracy compared to a direct measurement\cite{Kulikov2020}.

\begin{figure*}[t]
    \centering
    \begin{subfigure}{.45\textwidth}
        \centering
            \includegraphics[scale = 0.75]{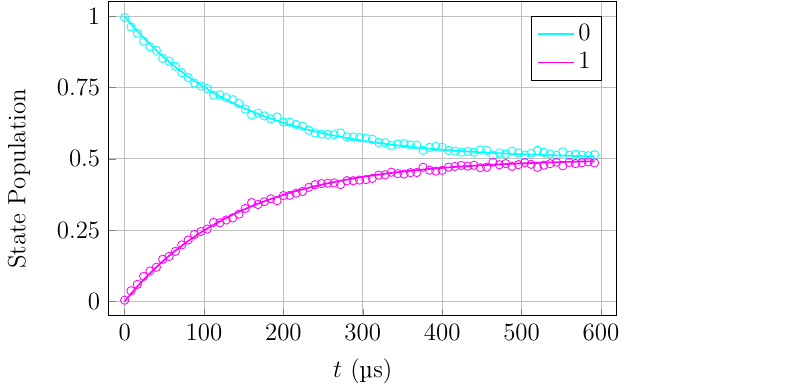}
        \caption{Optimised single-qubit $T_2$ sequence.}
        \label{subfig:t2e_1q_badfit}
    \end{subfigure}
    \begin{subfigure}{.45\textwidth}
        \centering
            \includegraphics[scale = 0.535]{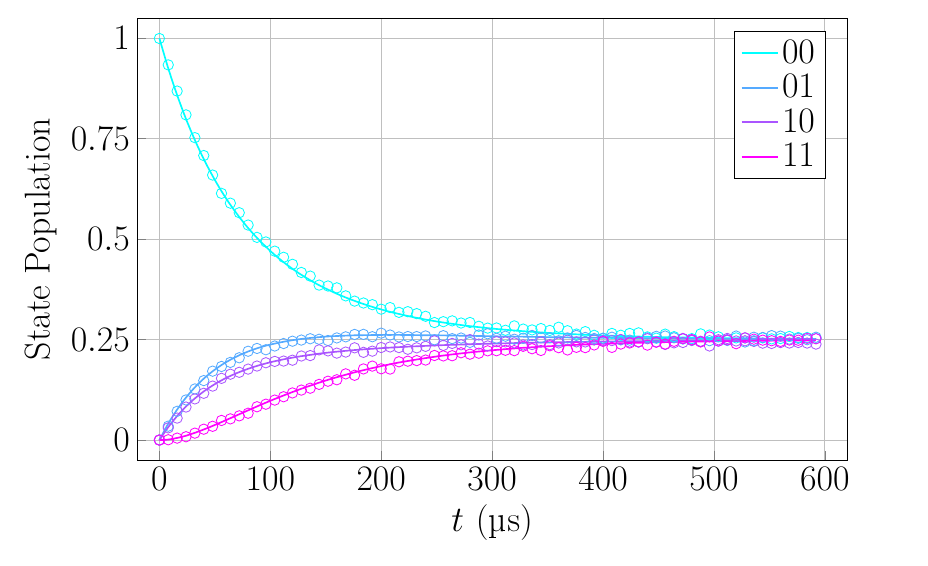}
        \caption{Optimised $T_2$ sequence for 2 qubits.}
        \label{subfig:t2e_fit}
    \end{subfigure}
    \begin{subfigure}{.45\textwidth}
        \centering
            \includegraphics[scale = 0.75]{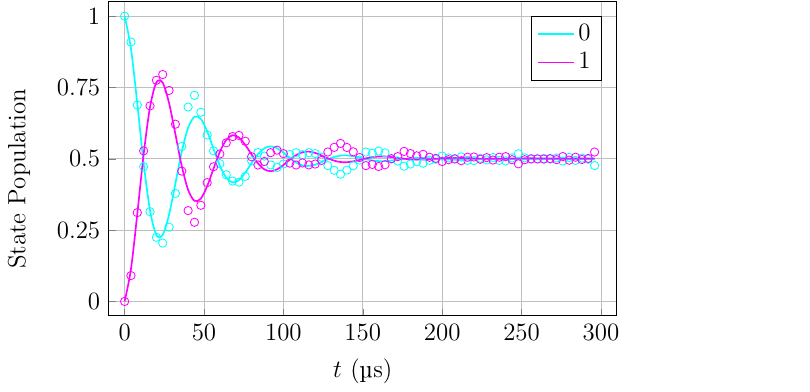}
        \caption{Optimised $T_2^*$ sequence for 1 qubit.}
        \label{subfig:t2s_fit}
    \end{subfigure}
    \begin{subfigure}{.45\textwidth}
        \centering
            \includegraphics[scale = 0.75]{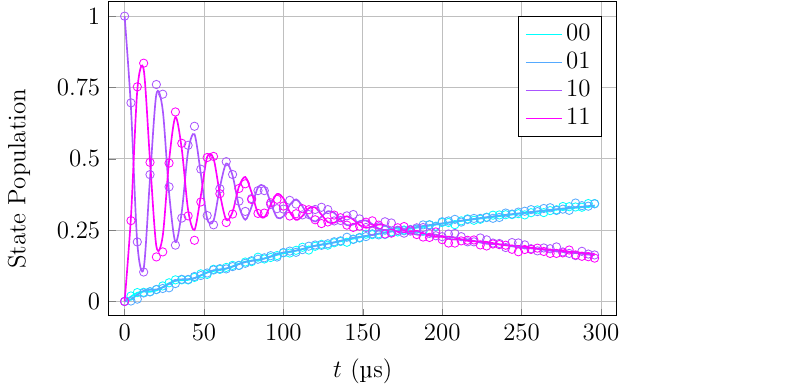}
        \caption{Optimised $T_2^*$ and $T_1$ sequences for 2 qubits.}
        \label{subfig:t2s_hx_fit}
    \end{subfigure}
    \caption{Optimised results for additional experimental circuits, including $T_2$, $T_2^*$, and combined procedures. The $T_2$ and $T_2^*$ times are characteristic quantum state decay times measuring the rate of dephasing from noise in the system. Experimental data represented by dots and numerical data for optimised parameters represented by solid lines. Results shown for single-qubit $T_2$ sequence (a), 2-qubit $T_2$ sequence (b), single-qubit $T_2^*$ sequence (c), and 2-qubit combination of $T_2^*$ and $T_1$ sequences (d).}
    \label{fig:extra_plots}
\end{figure*}

The $T_2$ sequence for 2-qubit subsystems demonstrated interesting and well behaved dynamics, as can be seen in \Cref{fig:extra_plots}b, showing a smooth exponential decay towards an evenly mixed state, which allowed for a direct demonstration of the Markovianity assumption with a more complicated architecture. This sequence produced results similar to those of the $T_1$ sequences, in terms of the hardware parameters that could be extracted, as they were consistent with expectations as well as the claimed parameters and optimised values of the $T_1$ sequences. The $T_2^*$ sequences proved to be slightly less reliable due to some off-resonance SPAM error, as discussed before, which would quickly lead to the results deteriorating in most cases and enforce the requirement of multiple attempts at the procedure to obtain accurate results. When these more accurate results were obtained, however, the optimisation process could work with great proficiency and give accurate fits to the observed data. The hardware parameters that this optimisation led to were not as would be expected by the trends set by previous sequences. Rather, all of the hardware parameters underwent a significant scaling phenomenon, shifting them all away from claimed parameters given by the backend providers. For example, the qubit frequencies underwent a dramatic shift by up to a factor of $2$, which is characteristic of the dephasing of the system, while also having average photon number contributions to the decay rates reflecting qubit temperatures on the scale of $1\,\si{\kelvin}$ rather than the expected $15\,\si{\milli\kelvin}$. Additionally, the actual decoherence time in the form of the decay rate $\gamma$ was much larger than the claimed parameters, which does not necessarily indicate a successful increase in qubit coherence but is likely rather a relic of the sequence possessing a more complicated form which would require more careful analysis to extract hardware information.

To help solve the associated difficulties of the $T_2^*$ sequence, the modified sequences shown in \Cref{fig:mixed_circuits}c and \Cref{fig:mixed_circuits}d, representing one qubit that undergoes the $T_2^*$ sequence with its neighbour who undergoes the $T_1$ and idle sequences, respectively. These modifications allowed for the damping of the resonant noise of 2 qubits being susceptible to SPAM errors, and provided a more granular look at the qubit dynamics. In the case of the circuit in \Cref{fig:mixed_circuits}d, this problem of resonant external noise remained persistent, although it showed a scaled intensity of errors, reinforcing the suspicion that the $H$ gates are in particular more difficult to implement effectively. In the case of the circuit in \Cref{fig:mixed_circuits}c, however, a new and interesting form of the qubit dynamics was observed, in \Cref{fig:extra_plots}d, which coupled the simple exponential decay of the $T_1$ sequence with the damped oscillator form of the $T_2^*$ sequence, which allowed for proficient optimisation and extraction of hardware parameters that were consistent with the results from the separate sequences showing that the mixing of these procedures does not produce any phenomena greater than the combination of its parts.

As another example of subsystems combining in a linear and separable manner, the $T_1$ sequence is worth returning to for a focused analysis of the predictions made by different subsystems of overlapping qubits. For this case, the single-qubit results for individual neighbouring qubits, which provided very accurate and reliable behaviours, can be compared to the results obtained by 2-qubit subsystems comprising those individual neighbours, as well as separate 2-qubit subsystems which overlap. This analysis goes one step larger to the 3-qubit system as well for an additional comparison point of the predicted hardware parameters and behaviours. For an example of the specifics of this kind of procedure, the first 3 qubits (indexed as 0, 1, and 2) of a 5-qubit system were analysed and optimised to extract parameters which agreed accurately with the claimed values, within a small margin of error, which justifies the use of optimisation. Then two sets of 2-qubit subsystems, focused on qubits (0,1) and (1,2) respectively, went through the optimisation process to produce values comparable with the single-qubit case as well as the doubly predicted qubit 1 parameters. These 3 qubits comprised the final overall system to investigate the scalability of this method.

The results that this procedure produced, as seen in \Cref{tab:1_qubit_results,,tab:2_qubit_results,,tab:3_qubit_results}, were very promising in that the single-qubit results demonstrated excellent accuracy compared to the observed data, while these results were replicated by the 2-qubit subsystems, which were internally consistent, and finally by the 3-qubit system to confirm this accurate extraction.

\begin{table*}[ht]
	\centering
	\begin{tabular}{|c|c|c|c|c|c|}
		\hline
		& \multicolumn{2}{c|}{IBMQ Reported} & \multicolumn{3}{c|}{Observed} \\
            \hline
		& $\omega \left( \si{\giga\hertz} \right)$ & $T_1 \left( \si{\micro\second} \right)$ & $\omega \left( \si{\giga\hertz} \right)$ & $T_1 \left( \si{\micro\second} \right)$ & $T \left( \si{\milli\kelvin} \right)$ \\
        \hline
		$q_0$ & $31.42$ & $100.24$ & $31.42 \pm 0.015$ & $101.23 \pm 0.12$ & $47.96 \pm 2.79$ \\
        \hline
		$q_1$ & $30.47$ & $106.95$ & $30.47 \pm 0.012$ & $108.31 \pm 0.19$ & $54.20 \pm 3.27$ \\
        \hline
		$q_2$ & $30.05$ & $101.45$ & $30.05 \pm 0.017$ & $105.92 \pm 0.21$ & $50.30 \pm 5.34$ \\
		\hline
	\end{tabular}
	\caption{IBMQ reported and observed hardware parameters for the single-qubit procedures to measure the qubit frequency, $\omega$, relaxation time, $T_1$, and qubit temperature, $T$. Error margin obtained through randomly sampled noise added to optimisation ansatz.}
	\label{tab:1_qubit_results}
\end{table*}

\begin{table*}[ht]
	\centering
	\begin{tabular}{|c|c|c|c|c|c|c|c|}
		\hline
		& \multicolumn{3}{c|}{IBMQ Reported} & \multicolumn{4}{c|}{Observed} \\
		\hline
		& $\omega \left( \si{\giga\hertz} \right)$ & $T_1 \left( \si{\micro\second} \right)$ & $J \left( \si{\mega\hertz} \right)$ & $\omega \left( \si{\giga\hertz} \right)$ & $T_1 \left( \si{\micro\second} \right)$ & $J \left( \si{\mega\hertz} \right)$ & $T \left( \si{\milli\kelvin} \right)$ \\
		\hline
		$q_{01,0}$ & $31.42$ & $100.24$ & \multirow{2}{*}{$8.31$} & $31.42 \pm 0.015$ & $102.43 \pm 0.23$ & \multirow{2}{*}{$5.87 \pm 1.24$} & $49.83 \pm 2.46$ \\
		\cline{1-3} \cline{5-6} \cline{8-8}
		$q_{01,1}$ & $30.47$ & $106.95$ &  & $30.47 \pm 0.014$ & $109.62 \pm 0.27$ &  & $65.55 \pm 3.65$ \\
		\hline
		$q_{12,1}$ & $30.47$ & $106.95$ & \multirow{2}{*}{$7.42$} & $30.47 \pm 0.014$ & $109.26 \pm 0.27$ & \multirow{2}{*}{$5.25 \pm 1.34$} & $67.63 \pm 3.15$ \\
		\cline{1-3} \cline{5-6} \cline{8-8}
		$q_{12,2}$ & $30.05$ & $101.45$ &  & $30.05 \pm 0.015$ & $108.31 \pm 0.26$ &  & $56.42 \pm 4.01$ \\
		\hline
	\end{tabular}
	\caption{IBMQ reported and observed hardware parameters for the 2-qubit procedures to measure the qubit frequency, $\omega$, relaxation time, $T_1$, inter-qubit coupling, $J$, and qubit temperature, $T$. Error margin obtained through randomly sampled noise added to optimisation ansatz.}
	\label{tab:2_qubit_results}
\end{table*}

\begin{table*}[ht]
	\centering
	\begin{tabular}{|c|c|c|c|c|c|c|c|}
		\hline
		& \multicolumn{3}{c|}{IBMQ Reported} & \multicolumn{4}{c|}{Observed} \\
		\hline
		& $\omega \left( \si{\giga\hertz} \right)$ & $T_1 \left( \si{\micro\second} \right)$ & $J \left( \si{\mega\hertz} \right)$ & $\omega \left( \si{\giga\hertz} \right)$ & $T_1 \left( \si{\micro\second} \right)$ & $J \left( \si{\mega\hertz} \right)$ & $T \left( \si{\milli\kelvin} \right)$ \\
		\hline
		$q_0$ & $31.42$ & $100.24$ & $8.31$ & $31.42 \pm 0.015$ & $101.16 \pm 0.11$ & $8.31 \pm 0.22$ & $50.59 \pm 3.16$ \\
		\hline
		$q_1$ & $30.47$ & $106.95$ & $-$ & $30.47 \pm 0.013$ & $107.77 \pm 0.22$ & $-$ & $67.92 \pm 3.24$ \\
		\hline
		$q_2$ & $30.05$ & $101.45$ & $7.42$ & $30.05 \pm 0.015$ & $108.08 \pm 0.18$ & $7.42 \pm 0.19$ & $66.37 \pm 4.21$ \\
		\hline
	\end{tabular}
	\caption{IBMQ reported and observed hardware parameters for the 3-qubit procedures to measure the qubit frequency, $\omega$, relaxation time, $T_1$, inter-qubit coupling, $J$, and qubit temperature, $T$. Error margin obtained through randomly sampled noise added to optimisation ansatz.}
	\label{tab:3_qubit_results}
\end{table*}

Additionally, to verify the capabilities of this method in finding the correct parameters to fit an accurate model, it cannot be heavily dependent on the initial parameters. Using purely random values may lead to local minima or exploding/vanishing gradients in the optimisation landscape\cite{Schuld2018,Cerezo2021}, so the values used were rather the claimed calibration parameters to ensure a relatively smooth optimisation path. However, such a simple method runs the risk of a lucky optimisation that gives ideal results while ignoring the actual optimisation landscape. To mitigate this concern, it is necessary to add noise to the initial parameter set and run the optimisation algorithm multiple times for statistical rigour\cite{Neelakantan2015}. This gives more reliable average results at the end of the process, with variance metrics to reinforce the accuracy of the results. The noise added to the parameters was sampled from a Gaussian PDF ($\mu = 0$, $\sigma = 1$) and run for 100 iterations. The standard deviation results are included in the numerical results presented in \Cref{tab:1_qubit_results,,tab:2_qubit_results,,tab:3_qubit_results}.

\begin{figure}[b]
	\begin{center}
		\includegraphics{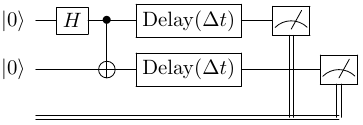}
	\end{center}
	\caption{Quantum circuit for the Bell state initialisation and decay. This circuit is constructed to initialise the maximally entangled 2-qubit Bell state, followed by the quantum state decoherence to monitor the influence of quantum noise.}
	\label{fig:bell_circuit}
\end{figure}

These results not only reflect the findings of the $T_1$ sequences discussed previously, about the generalised form of the Hamiltonian and more accurate decay parameters, but also confirm the viability of this method to predict larger systems of qubits. This means that multiple 2-qubit systems can be run and analysed, similarly to tomography procedures, and then used to extract parameters and act as a form of calibration. The 2-qubit results can also be combined through tensor products to obtain a tomography of a larger system which would ordinarily grow exponentially in difficulty.

\begin{figure*}[t]
    \centering
    \begin{subfigure}{.45\textwidth}
        \centering
            \includegraphics[scale = 0.55]{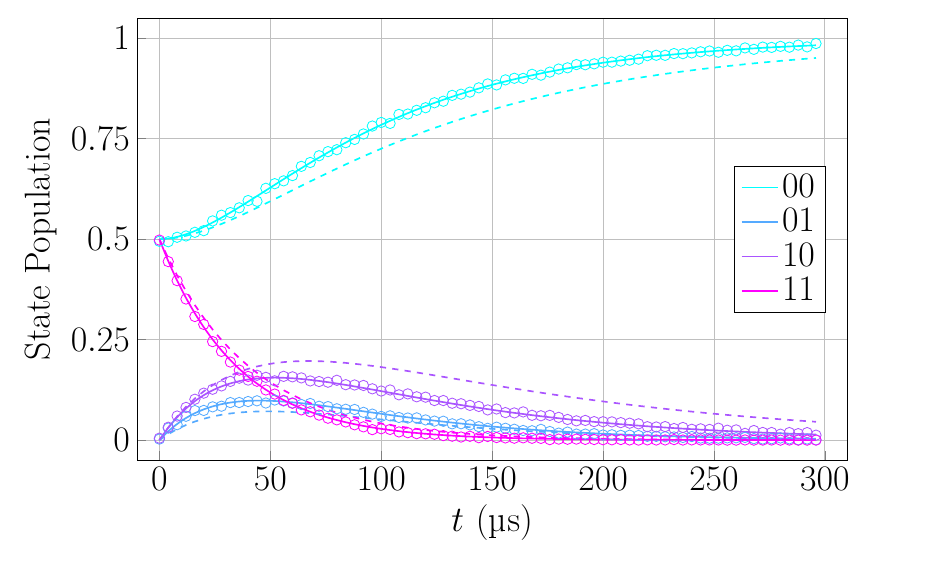}
        \caption{Predicted Bell state decay sequence.}
        \label{subfig:bell}
    \end{subfigure}
    \begin{subfigure}{.45\textwidth}
        \centering
            \includegraphics[scale = 0.55]{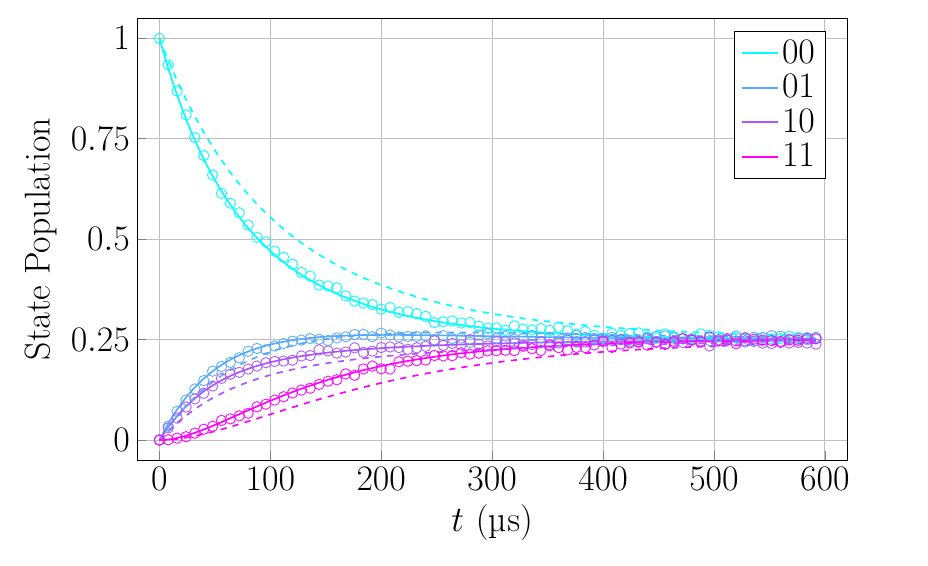}
        \caption{Predicted $T_2$ sequence for 2 qubits.}
        \label{subfig:t2e}
    \end{subfigure}
    \caption{Optimised results for predicted behaviours of circuits based on 2-qubit $T_1$ parameter extraction. Experimental data represented by dots, numerical data for optimised parameters represented by solid lines, calibration ansatz expected behaviour represented by dashed lines. Results shown for 2-qubit Bell state decay, from \Cref{fig:bell_circuit} (a), 2-qubit $T_2$ sequence (b).}
    \label{fig:predictions}
\end{figure*}

This method is, however, fairly unreliable as a self-contained single-run tomography procedure which is mostly the fault of the backend rather than the method. The backend is hindered by the amount of traffic it faces with it being a cloud-based open-access service which occasionally leads to queue times lasting longer than several calibrations which occur hourly. This leads to some procedures being run on different calibration profiles and after different reset periods which affect the hardware parameters dictating the qubit dynamics, which finally leads to inconsistencies in the extracted optimal data, which destroys the integrity of the tomography. Additionally, as the backend is open-access, the attainable information is fairly limited, for example with the device temperatures, which would be able to be monitored more efficiently with direct access to the devices. Although this method is meant to be useful especially in such cases to extract information which is not being actively monitored, there is a significant improvement to be made in the feedback from the devices which are ultimately controlled by the backend compilation which provides a larger margin for error between the gate composition and the execution of the quantum circuits.

Despite the difficulties faced in this setting of the demonstrational process, the data show very promising results for the viability of this as a form of tomography to probe the dynamics of larger systems of connected qubits in NISQ devices. For isolated devices with direct control within a laboratory\cite{Werninghaus2021}, this method could provide significant insight into unaccounted-for noise sources deteriorating the qubit coherence\cite{Vepsaelaeinen2020,Martinis2021}, and for open-access devices\cite{GarciaPerez2020} this method can serve as an additional calibration method to obtain the most current hardware parameters for higher precision in the simulation of experiments such as quantum chemistry or open quantum systems.

Finally, this framework also offers the major advantage of predictive capabilities. The methodology can be applied to a simple $T_1$ sequence to extract the hardware parameters in the simplest case, and then immediately be used reliably as a replacement for other sequences. For example, the $T_1$ sequence can be run and optimised, and then the $T_2$ parameter can be derived from this without the need for another circuit run. This is because the output of the framework is independent of initial conditions or circuit structure. As can be seen in \Cref{fig:predictions}, the parameters extracted from the 2-qubit $T_1$ sequence can be directly applied to the calculation of the Bell state (\Cref{fig:bell_circuit}) and $T_2$ evolution (\Cref{fig:basic_circuits}c) to produce a significantly more accurate fit than the calibration parameter ansatz. So not only does the method extract all relevant parameters for the benchmark circuit being run, but also for all other circuits which can be executed on the hardware. This makes the framework stand out as a significant improvement over standard tomography and calibration methods, which only extract the directly relevant information in narrowly defined procedures. This significantly improves reliability and runtime speedup of noise modelling in quantum devices.

\section*{Conclusion}
\label{sec:Conclusion}

In this work the fundamentals of several fields of research, such as quantum computing, superconducting qubit hardware, and open quantum systems, were introduced and discussed to lay a foundation for the understanding of the topics and their relevance to each other. This relevance was leveraged into a discussion of the overlap of these topics in the realm of quantum noise and error mitigation, which is covered primarily by current research. Recent research has shown that there is an uncertainty about the exact behaviour of noise and dissipation in NISQ transmon devices. The investigation of this uncertainty was the basis of this work, upon which a new method for parameter extraction and re-calibration was provided.

This new framework is a method of using basic benchmarking procedures commonly used to measure the relaxation and decoherence times, $T_1$ and $T_2$, of quantum states, and numerically replicated through a Markovian quantum master equation to measure how well the claimed hardware parameters fit the observations. This process was expanded to use optimisation methods to improve the agreement between numerical results and observed data to extract more accurate values of the hardware parameters which dictate the dynamics of the system.

This gave rise to interesting phenomena, such as more generalised Hamiltonians containing broader descriptions of qubit frequency and coupling, as well as proving the proficiency of this method to extract all of the hardware parameters without needing to revert to various specialised experiments which measure the parameters individually. The consistency of this method was verified through using various sizes of qubit subsystems which each produced values for hardware parameters for individual qubits, which were congruent to a reliable degree of accuracy that the process could be seen as successful.

This method serves as a proof of concept of subsystem Markovian tomography, which can be stitched together to provide a mapping of qubit dynamics in larger systems without the exponential growth in difficulty of full quantum tomography.

More improvements can be made to this method in future works, such as assisting the performance with machine learning methods\cite{Baum2021,Naicker2022} and performing more accurate calibration and tomography to mitigate any external errors which are unavoidable in open-access NISQ devices\cite{Wittler2021,Helsen2022}, or finding mechanisms to improve qubit coherence times. Additional improvements which are currently being researched are the inclusions of more comprehensive noise models which account for phenomena such as cross-talk\cite{Zhou2023}. This work also opens a new path for further investigation into more generalised models, which can uncover finer details that are typically suppressed in approximation, such as qubit frequency vectors and coupling tensors, but could prove crucial in further quantum engineering in the path to fault-tolerance.

\bibliography{main}

\section*{Acknowledgements}

This work is supported by the South African Research Chair Initiative, Grant No. 64812 of the Department of Science and Innovation and the National Research Foundation of the Republic of South Africa. Support from the NICIS (National Integrated Cyber Infrastructure System) e-research grant QICSI7 is kindly acknowledged. We acknowledge the use of IBM Quantum services for this work. The views expressed are those of the authors, and do not reflect the official policy or position of IBM or the IBM Quantum team. This work was done primarily at the University of KwaZulu-Natal, Durban, which is  kindly acknowledged.

\section*{Author contributions}

D.B. and I.S. designed the experiments; D.B. performed the experiments, data analysis, and wrote the code; I.S. conceived the research idea; F.P. supervised the project; D.B. wrote the manuscript; All authors reviewed and proofread the manuscript.

\section*{Data availability}

The data presented in this paper was created using Python, particularly the Qiskit and JAX packages. The code created to obtain and analyse these results is available on GitHub at \url{https://github.com/deanbrand/MarkovianModelling\_Qiskit}.

\section*{Competing interests}

The authors declare no competing interests.

\section*{Additional information}

\textbf{Correspondence} and requests for materials should be addressed to D.B.

\end{document}